\documentclass[apj]{emulateapj}

\shorttitle{Space Velocities of Globular Clusters}
\begin{document}
\title{Space Velocities of Southern Globular Clusters VII. NGC 6397, NGC 6626 (M 28) and NGC 6656 (M 22) }

\author{Dana I. Casetti-Dinescu\altaffilmark{1}, 
Terrence M. Girard\altaffilmark{1}, 
Lucie J\'{i}lkov\'{a}\altaffilmark{2}, 
William F. van Altena\altaffilmark{1},
Federico Podest\'{a}\altaffilmark{3} and
Carlos E. L\'{o}pez\altaffilmark{3}
}

\altaffiltext{1}{Astronomy Department, Yale University, P.O. Box 208101,
New Haven, CT 06520-8101, USA, dana.casetti@yale.edu, terry.girard@yale.edu, william.vanaltena@yale.edu}
\altaffiltext{2}{Department of Theoretical Physics and Astrophysics, 
Faculty of Science, Masaryk University; address: Kotl\'a\v{r}sk\'a 2, 
CZ-61137 Brno, Czech Republic, jilkoval@physics.muni.cz}
\altaffiltext{3}{Universidad National de San Juan, Observatorio Astron\'{o}mico ``F\'{e}lix Aguilar'' and
Yale Southern Observatory, Chimbas, 5413 San Juan, Argentina}

\begin{abstract}
We have measured the absolute proper motions of globular clusters NGC 6397, NGC 6626 (M 22) and NGC 6656 (M 28)
as part of our ongoing Southern Proper-Motion Program. The reference system is the ICRS via 
$Hipparcos$ stars for these three low galactic latitude clusters. Formal errors range between
$\sim 0.3$ and $0.7$ mas yr$^{-1}$. Notable is the result for NGC 6397 which 
differs by 2.5 mas yr$^{-1}$ from two HST determinations, while agreeing with previous ground-based ones.
We determine orbits for all three clusters in an 
axisymmetric and barred model of the Galaxy and discuss these in the context
of globular-cluster formation. M 22 is a well-known cluster with an iron abundance spread;
such clusters are now believed to have formed in massive parent systems that can retain
ejecta of core-collapsed SNe. We find that the five currently-accepted globular clusters 
with iron/calcium abundance spread show
orbits unrelated to each other, thus suggesting at least five independent, massive progenitors that 
have contributed to the build-up of the Milky-Way halo. 
\end{abstract}


\section{Introduction}

Over the past few years a new view for the formation of globular clusters and their 
contribution to the Milky-Way (MW) halo has emerged. Photometric and spectroscopic studies
have established that globular clusters are not  simple stellar populations, but rather are formed from two or more
generations of stars with the latter generation(s) formed from chemically enriched material 
from the previous generation (see e.g., review by Gratton et al. 2012).
That these systems were able to retain ejecta from massive AGB stars, including ejecta of
core-collapsed SNe for a handful of clusters, to form subsequent generations of stars 
implies large progenitor masses (e.g., Joo \& Lee 2013, Gratton et al. 2012 and references therein,
Valcarce \& Catelan 2011). Through subsequent dynamical evolution it is suggested globular clusters lose 
large amounts of mass contributing to the MW halo. Martell et al. (2011) estimate that a minimum fraction of $17\%$ 
of the halo is from globular clusters while Gratton et al. (2012) suggest a fraction $> 50\%$;
this is remarkable, since the present-day luminous mass of the MW globular clusters is only about $\sim 1 \%$  that of the halo.
In this new light, globular clusters appear to be important players in the formation of the MW halo, and thus
a better understanding of these systems and their evolution is worthwhile. 

Here, we continue our long-term efforts to determine absolute proper motions 
of globular clusters and derive their orbits in a realistic MW potential. Such data
help with the quantifying of their tidal disruption, and with searching for possible dynamical associations
between clusters and dwarf spheroidals, in an effort to understand their origin.
We present results for three clusters, NGC 6397, NGC 6626 (M 28) and NGC 6656 (M 22), which have 
previous absolute proper-motion determinations. Our new results indicate disagreement between
ground-based and HST-based measurements for NGC 6397, imply a new type of orbit for NGC 6626 than previoulsy
derived, and provide a higher accuracy measure for NGC 6656.
Ground-based measurements for these clusters were presented
in the work by Cudworth \& Hanson (1993, hereafter CH93, and references therein). HST-based measurements
were done for NGC 6397 (Kalirai et al. 2007, Milone et al. 2006) and for NGC 6626 (Chen et al. 2004).
Since all three clusters are at low Galactic latitude, the question of the inertial reference system
poses some problems. Our work, which is calibrated on the ICRS via $Hipparcos$ stars, provides 
an important comparison.
NGC 6656 (M 22) is an extended blue horizontal branch (EHB) cluster (e.g., Lee et al. 2007) with a known
iron abundance spread (Da Costa et al. 2009, Marino et al. 2009, 2011) presenting the possibility that it might 
be the nuclear cluster of a formerly massive system 
(like M 54 in the Sagittarius dwarf galaxy,
e.g., Monaco et al. 2005). NGC 6626 is also a moderate EHB cluster as classified by Lee et al. (2007), 
while NGC 6397 is a less massive, more traditional globular cluster.
We adopt various parameters for these clusters from the database of 
Harris (1996, 2010 updated and to be referred to as H96).

In the following Section we describe our measurements, while the proper-motion determinations are
outlined in Section 3. In Section 4 we compare our results with existing measurements, and
in Sections 5 and 6 we give the velocities and the orbits determined 
in an axisymmetric and bar model,
respectively. In Section 7 we present a discussion of our results.

\section{Measurements}
Our program to measure absolute proper motions of globular clusters
is a distinct part of the Southern Proper Motion Program (SPM) which 
was described in a series of papers: Platais et al. (1998), 
Girard et al. (1998, 2004, 2011). 
The cluster program of the SPM has provided measurements
for 34 clusters reported in a series of papers: Dinescu et al. (1997, 1999ab, 2003) and 
Casetti-Dinescu et al. (2007, 2010), named Papers I through VI, respectively.
Including the sample presented here, there are 
37 clusters measured by the SPM, or $59\%$ of the total number of clusters 
with such measurements\footnote{An updated version of the catalog with all globular clusters with absolute 
proper motions can be found at www.astro.yale.edu/dana/gc.html}.

For convenience, we describe here in brief the major aspects of the program and 
measurements extracted from a more complete description given in Paper VI.
The first epoch of the survey is entirely photographic, and
was taken from 1965 to 1979.
The second epoch is approximately one third photographic
(taken from 1988 to 1998) and two thirds CCD-based (taken from 2003 through 2008). For the clusters, we also take frames centered on each cluster in addition 
to the ``brick'' coverage of the main survey.
We list the SPM field centers, the photographic and CCD material used in the reductions, and 
the epochs of the observations in Table 1. All three fields lacked complete CCD coverage from our 
main SPM survey; these gaps are at the edges of the fields, and amount to between 5 and 17$\%$ of the 
total SPM field area, as listed in Table 1. For these areas, second-epoch positions are taken from $Hipparcos$, 
Tycho-2 (Perryman et al. 1997), UCAC2 (Zacharias et al. 2000) and 2MASS 
point source catalog (Skrutskie et al. 2006). Given the 
construction of the input list of objects to be measured, these positions are 
predominantly from 2MASS.

The clusters presented here are at low Galactic latitude and toward the bulge direction. Since 
extinction is high in these directions, we lack extragalactic objects
to tie to an inertial reference frame; therefore, we use 
$Hipparcos$ stars to correct to absolute proper motions.
Thus, the proper motions are on the 
International Celestial Reference System via $Hipparcos$. 
Only in the case of NGC 6397 have we found a handful of 
galaxies to check our $Hipparcos$-based zero point.

\begin{table*}[htb]
\caption{SPM Field Centers and Observational Material}
\begin{tabular}{ccrrrrllr}
\tableline
\\
\multicolumn{1}{c}{NGC} &
\multicolumn{1}{c}{SPM} & 
\multicolumn{1}{c}{R.A.} &
\multicolumn{1}{c}{Dec.} &
\multicolumn{1}{c}{l} &
\multicolumn{1}{c}{b} &
\multicolumn{1}{l}{Photographic} &
\multicolumn{1}{l}{SPM CCD$^{a}$} &
\multicolumn{1}{c}{Other$^{b}$} \\
& & \multicolumn{2}{c}{(B1950)} & \multicolumn{2}{c}{($\arcdeg$)} & \multicolumn{1}{l}{Pairs~~~Epoch} &
\multicolumn{1}{l}{$N_{S}~~N_{C}~~~$Epoch} & 
\multicolumn{1}{c}{(area)} \\
\tableline
\\
       6397 & 191 & 264 & -55 & 337 & -13 & 1(BY)$^{c}$ 1968.6 & $107~~9~~2005-6, 2011$ & 5\% \\
       6626 & 582 & 275 & -25 &   8 &  -5 & 2(Y)$^{d}$~~1967.4, 1969.5 & $240~~9~~2003, 2007-8,2011$ & 17\% \\
       6656 & 583 & 280 & -25 &  10 &  -9 & 2(BY) 1967.4, 1978.7 & $167~~9~~2007-8, 2011$ & 15\% \\
\tableline 
\multicolumn{8}{l}{{$^a~$}Number of frames taken by the main survey ($N_S$) and on clusters ($N_C$).} \\
\multicolumn{8}{l}{{$^b~$}2nd epoch positions from TYCHO2, UCAC2, or 2MASS for \% area of the SPM field.} \\
\multicolumn{8}{l}{{$^c~$}B refers to blue plates, while Y to visual plates - see Section 2.1.} \\
\multicolumn{8}{l}{{$^d~$}although B plates exist for this field, they were not used due to severe 
magnitude equation.}
\end{tabular}
\end{table*}

\subsection{Photographic Measurements}

The SPM plates were taken with the 51-cm double astrograph at Cesco Observatory in two passbands:
blue (103a-O, no filter) and visual (103a-G + OG515 filter). The plate scale is 55.1''/mm, 
and each field covers $6.3\arcdeg \times 6.3\arcdeg$. The plates contain two exposures: 
one of 2 hours that reaches $V\sim 18$ and an offset of 2 minutes. During both exposures, 
an objective grating is used which produces a series of diffraction images on 
either side of the central, zero-order images. The multiple sets of images for bright stars (for $V \ge 14$, there are only long exposure,
zero-order images) allow us to detect and model magnitude-dependent systematics that 
affect positions and consequently proper motions.
The method has been described and thoroughly tested by Girard et al. (1998), and it shows
that $Hipparcos$ stars with $V \sim 9$ and faint cluster stars ($V \ge 14$) can be placed on
a system that is largely free of systematics.
The plates are measured with the Yale PDS microdensitometer in an object-by-object mode with
a pixel size of 12.7 $\mu$m. For each SPM field, we measure a preselected set of
stars (see also Papers IV, V and VI). This set consists of all $Hipparcos$ stars 
(typically 100 per field), a set of 
$\sim 1800$ Tycho-2 stars (Perryman et al. 1997), another set of $\sim 1000$ bright 
($R_{UCAC2} < 13.5$) UCAC2 stars, 
reference stars and cluster-region stars. Reference and cluster stars are 
selected from UCAC2 and 2MASS catalogs. Cluster-region stars 
are selected to reside within a $7\arcmin$ 
radius from the cluster center. Reference stars consist of two subsets: one set uniformly 
distributed over the entire field, the other as a ring around each cluster with radius  
$7\arcmin \le r \le 11\arcmin$. Within the ring and cluster area, all objects found in 
UCAC2 or 2MASS are selected for the input list.
For the uniformly distributed reference stars over the 
entire field, we select a fraction of UCAC2 and 2MASS stars imposed by the time limitation of the PDS scans.
The bright stars in the input catalog assure an appropriate magnitude range of various diffraction 
orders with which to model magnitude-dependent systematics. Only $Hipparcos$ stars are 
used to determine the correction to absolute proper motions.
The reference stars (ring and entire field) are used to map plate positions into one another 
as well as CCD positions into the photographic positions.
During these long PDS scans (up to 24 hours) we monitor a set of 8 to 10 stars that are used to correct for
thermal drift. Positions, instrumental magnitudes and other image parameters are
derived with the Yale centering routines that fit a two-dimensional elliptical Gaussian function to 
the image profile (Lee \& van Altena 1983).

\subsection{CCD Measurements}
The CCD system mounted on the double astrograph consists of two main cameras: a 4K x 4K 
PixelVision (PV) camera (15 micron pixel) in the focal plane of the yellow lens and
an Apogee Alta 2K x 2K (12 micron pixel) camera in the focal plane of the blue lens.
The PV's field of view is
$0.93\arcdeg \times 0.93\arcdeg$ with a scale of 0.83''/pix.
The Apogee Alta camera covers
a $0.38\arcdeg \times 0.38\arcdeg$ area with a scale of 0.74''/pix. The PV data were used for 
both astrometry and $V$ photometry, while the Apogee data were used only for $B$ photometry.
The observations are taken with the diffraction grating oriented at $45\arcdeg$. This  
ensures a dynamical range of 10 mag (i.e., 6 from the CCD plus 4 from the grating).
Exposure times are 120 sec, reaching the same depth as the first-epoch plates.
For each cluster nine frames are taken, in addition to the survey CCD frames taken 
for each SPM field (Tab. 1). These on-cluster frames have exposures of 30, 60 and 120 sec.
The astrometric reduction of the CCD data is thoroughly described in Papers V and VI, and
in Girard et al. (2011). Thus, here we only summarize the procedure applied.
The CCD data are corrected for bias, dark (when needed) and flat fielding. 
The SExtractor code (Bertin \& Arnouts 1996) is used for object detections, 
aperture photometry and preliminary x,y centers. Final x,y centers are derived by fitting
two-dimensional elliptical Gaussian functions to image intensities. The obtained positional precision  
is $\sim 20$ mas per single measurement, for $V=7-15$.
For each CCD frame, we apply two corrections before merging them into a 
``pseudo'' plate. The first correction is the optical field angle distortion (OFAD)
which has been determined from stacked residuals into UCAC2, and is applied as a correction map
 to the CCD data 
(see Paper V and Girard et al. 2011)\footnote{UCAC2 is adequate for the purposes of this work i.e., selecting reference stars and defining the distortion mask. At any rate,  at the time we were contructiong the mask, UCAC3 was problematic, while UCAC4 was not available.}. The second correction refers to the positions 
of different grating-order 
images that must be placed
on a common system with those of the central-order images.  We found offsets between the 
average position of symmetric image orders and the position of the central order. These offsets 
are not related to the magnitude of the object, as is the case for photographic plates. 
Therefore these offsets were corrected as such, a simple offset between different  image 
order systems (see also Paper V, Girard et al. 2011).
Next, all of the CCD frames for a given SPM field from the main survey are linked 
to produce a $6\arcdeg\times6\arcdeg$ CCD pseudo-plate. The procedure to link the CCD frames was 
developed for the construction of the SPM4 catalog and is
described by Girard et al. (2011).  In brief, the
CCD x,y positions of each frame are transformed
onto the system of the UCAC2 to facilitate pasting together the 100-200 frames 
that comprise a $6\arcdeg\times6\arcdeg$ SPM field.  An iterative overlap method is 
employed to
perform this task using Tycho2 stars as an external reference system to
ensure that systematics from the individual frame overlaps do not
accumulate. The CCD frames that contribute to any one field's pseudo-plate
may actually span two to three years' range in epoch whereas an actual plate
has a single epoch.  For this reason,
once preliminary proper motions are determined from a second-epoch 
pseudo-plate, an improved pseudo-plate is constructed, one that incorporates
the preliminary proper motions prior to pasting the frames together in a
new overlap solution. 

The CCD $BV$ photometry is derived from aperture photometry calibrated into 
Tycho2 (corrected to the Johnson system), and other external catalogs
such as the Guide Star Photometric Catalog II (Bucciarelli et al. 2001).
Photographic photometry from the first-epoch plates calibrated on Tycho2
is adopted when no astrograph CCD photometry is available,
(see Girard et al. 2011).

\section{Proper Motions}

Proper motions are determined using the central-plate overlap method (e.g., Girard et al. 1989).
We transform all measurements --- CCD pseudo-plate, on-cluster CCD frames and 
photographic positions of various grating-order systems --- into one master 
photographic position system, that of a visual plate's central diffraction-order images.
This transformation is modeled as a polynomial function of x,y coordinates of up to 5th order.
Stars with 
magnitudes between $V = 13$ and 18 distributed uniformly over the field and with a higher density
in a ring around each cluster are used as reference stars to model the transformation. 
The number of reference stars ranges between $\sim 200$ 
and 7000 stars. 
In the first round of plate transformations, stars are assigned  zero proper motions.
The process is repeated, iteratively, with proper motions updated at each iteration.
A linear least-squares fit of positions as a function of time gives the proper motion of each object.
Positional measurements that differ by more than $0.2\arcsec$ from the best fit line are rejected as outliers.
The proper-motion error is given by the scatter around the best-fit line.
Throughout this paper, proper-motion units are mas yr$^{-1}$.
Also, $\mu_{\alpha}^* = \mu_{\alpha}$cos$~\delta$.

\subsection{Absolute Proper Motions}

The correction to absolute proper motion is determined from the difference between
the $Hipparcos$ proper motions and the relative proper motions obtained in the previous Section. 
These differences are also used 
to check for various trends as a function of magnitude and position on the plate.
During the analysis, we found that the B plates in SPM field 582 have very strong magnitude 
equation which our correction scheme could not correct at a reliable level.
Therefore, for this field, we used only the Y plates (see Tab. 1).

The proper-motion differences $\Delta \mu = (\mu_{Hipparcos} - \mu_{SPM})$ as a function of
$V$ magnitude are shown in Figure 1 for each SPM field. Average formal proper-motion errors 
of individual $Hipparcos$ stars are shown in the bottom right of each panel, and they range between
2.1 and 2.4 mas yr$^{-1}$. Fields 582 and 583 however, show a larger scatter than the
formal proper-motion errors. This is largely due to crowding in these low
Galactic latitude, toward the bulge fields. Also, in these fields there are 
proper-motion gradients across the field as shown in Figure 2.
Here, we present the vector point diagrams (VPDs) in the left panels, and the proper motions
in both coordinates as a function of X and Y coordinates (in mm at 55.1 ``/mm) in the 
middle and right panels. Fields 582 and 583 show gradients across the field which are likely due to 
true proper-motion variation. Both fields have large reddening ($E_{(B-V)} \sim 0.3 - 0.4$ at the cluster centers), 
which also indicates significant differential reddening across the field. 
Thus, the mean motion of the reference system may change across the field, in part due to sampling different mean brightness stars, 
in part due to true kinematical variation as one moves towards the bulge. 
The correction to absolute proper motion, or the zero point, is
determined from the average of the proper-motion differences as follows. First,
we eliminate stars brighter than $V = 5.0$, and with proper-motion differences larger than 30 mas yr$^{-1}$.
To determine the average and its error, we use probability plots (Hamaker 1978) trimmed at $10\%$ at each side,
to eliminate outliers. These values are listed in
Table 2, columns three and four, together with the number of $Hipparcos$ stars used in the determination.
They are also represented with a cross in Fig. 2 - left panels,
and a horizontal line in the middle and right panels. They are taken to represent the 
correction to absolute at the center of the SPM field. In Fig. 2, we also show the location of each 
cluster with a vertical bar. Thus, NGC 6397 and NGC 6626 have no field correction applied,
since for NGC 6397 no gradients are apparent, and NGC 6266 is located close to the field center.
For NGC 6656 we have applied a correction for field dependence in $\mu_{\alpha}^{*}$.
The final adopted values for the zero point for each cluster are listed in Table 2, columns
six and seven.

\begin{table*}[htb]
\caption{Corrections to Absolute Proper Motions}
\begin{tabular}{rrrrrrr}
\tableline
\multicolumn{1}{c}{NGC} &
\multicolumn{1}{c}{SPM} &
\multicolumn{1}{c}{$\Delta\mu_{\alpha}^*$ @ center} &
\multicolumn{1}{c}{$\Delta\mu_{\delta}$ @ center } &
\multicolumn{1}{c}{$N_{Hip}$}  &
\multicolumn{1}{c}{$\Delta\mu_{\alpha}^*$ @ cluster} &
\multicolumn{1}{c}{$\Delta\mu_{\delta}$ @ cluster } \\
 & & \multicolumn{1}{c}{(mas~yr$^{-1}$)} &
 \multicolumn{1}{c}{(mas~yr$^{-1}$)} &  & 
\multicolumn{1}{c}{(mas~yr$^{-1}$)} &
\multicolumn{1}{c}{(mas~yr$^{-1}$)} \\
\tableline
\\
6397 & 191 & -1.08(0.27) & -4.79(0.24)& 129 & -1.08(0.27) & -4.79(0.24) \\
6626 & 582 & -0.68(0.45) & -3.71(0.34)& 84  & -0.68(0.45) & -3.71(0.34) \\
6656 & 583 & -0.47(0.43) & -3.91(0.41)& 82  & 0.11(0.43) &  -3.91(0.41)\\
\tableline
\end{tabular}
\end{table*}

\begin{figure*}
\includegraphics[scale=0.9]{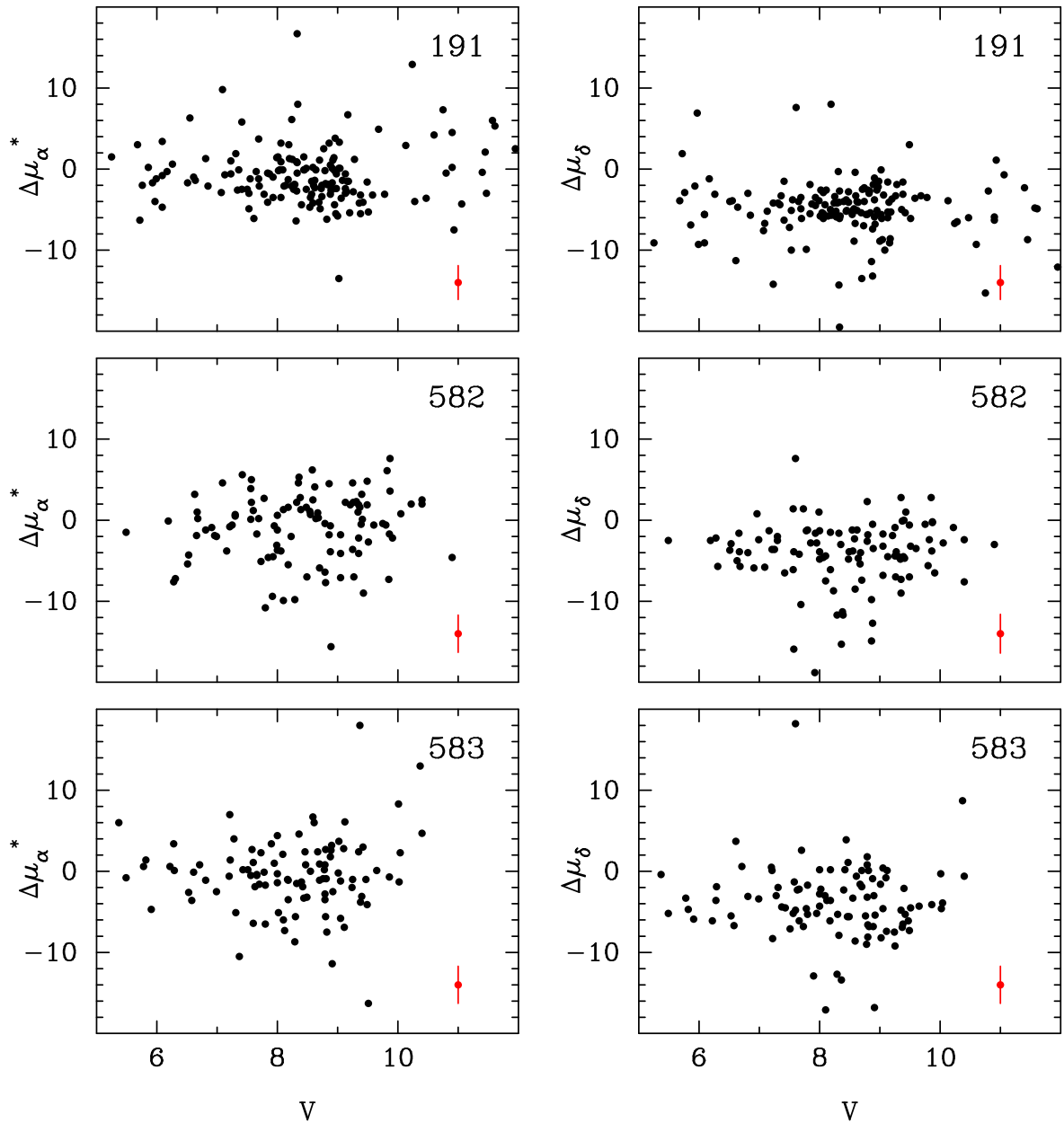}
\caption{Proper-motion differences (Hipparcos-SPM) as a function of magnitude for the three SPM fields as labeled. 
A typical error bar is shown in the lower right. Proper-motion units are mas yr$^{-1}$.}
\end{figure*}
\begin{figure*}
\includegraphics[angle=-90,scale=0.7]{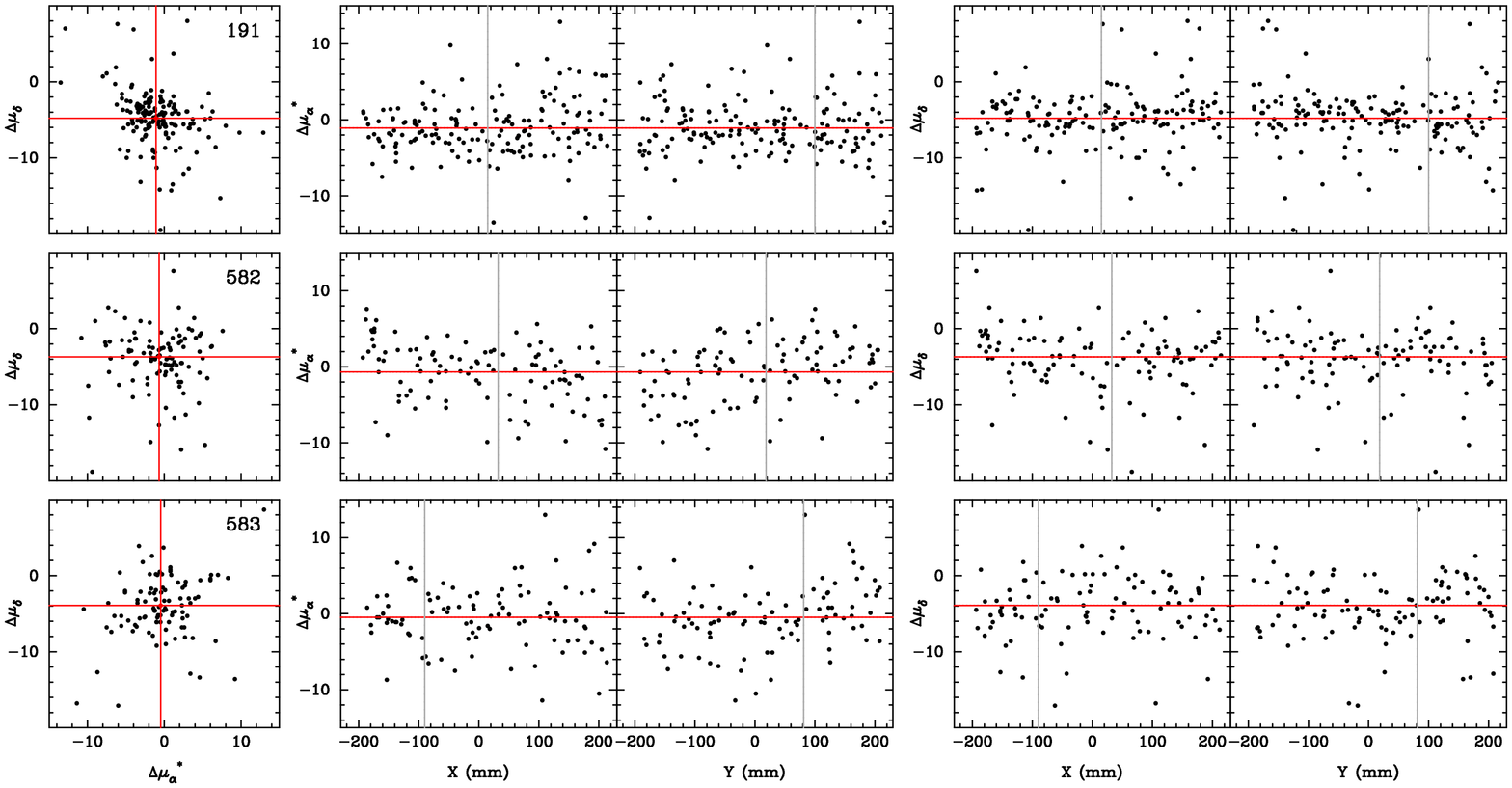}
\caption{(Hipparcos-SPM) proper-motion differences (left panels). These differences 
as a function of position on the plate are also shown (middle and right panels). The red 
lines represent the mean correction to absolute as determined in the text. The vertical lines 
show the position of the cluster on each field.}
\end{figure*}

\subsection{Cluster Proper Motions}

For NGC 6397, we obtain four estimates of the mean relative proper motion of the cluster with
respect to field stars; these are based on the way cluster members are selected. 
For the first two estimates, we select red giants (RGs) and blue horizontal branch (BHB) 
stars in the cluster from 2MASS photometry. 
Then we use proper-motion membership probabilities from the study by 
CH93 and references therein to earlier work of the same group.
CH93 used photographic plates of more appropriate
plate scale for proper-motion membership probabilities, but that go only to a limiting 
$V \sim 16.0$, and cover a small field of view. Throughout this paper we will refer to these
probabilities as $P_{\mu}^C$.
Lastly, we separate the cluster from the field using the
proper-motion distribution obtained in this study for all stars measured within $11\arcmin$
of the cluster center. For the samples defined by CH93 membership probabilities, 
or by 2MASS photometry we use probability plots trimmed at $10\%$ to estimate 
the average proper motion and its error. In Figure 3 we show the 2MASS color-magnitude diagram 
for stars within $11\arcmin$ of the cluster center (top-left panel). Here, the samples 
of RG and BHB stars are highlighted; also shown with star symbols are objects with 
CH93 proper-motion membership probability $P_{\mu}^C > 80\%$. The subsequent panels
show the VDPs of all these samples. The RG sample is the most contaminated by
field stars as there is some overlap with foreground/background stars in this region
of the CMD. Thus, to estimate the average motion for this sample, we first select
stars within 8 mas~yr$^{-1}$ from an initial estimate of the mean cluster motion
given by the peak of the proper-motion distribution. To this trimmed sample we then apply the 
probability plot estimate. The proper-motion values from these three estimates are
listed in Table 3, and indicated in Fig. 3.
In Figure 3 we also show the VPD of all stars within $11\arcmin$ of the cluster center,
and with $V \le 16.0$. The cluster is very well separated from the field, and thus we
can use a traditional method to fit the proper-motion distribution in each coordinate
with the sum of two Gaussians, one representing the field, the other the cluster
(see e.g., Girard et al. 1989). The observed proper-motion distribution is constructed 
from the data smoothed by individual proper-motion errors. The mean and the width of the fitted
Gaussian to the cluster sample represent the mean proper motion
of the cluster, and its error (when divided by the square root of the number of cluster 
stars in the fit).
This estimate is also listed in Table 3. The bottom-right panel of
Fig. 3 shows the proper-motion distribution in both coordinates: dot-dashed lines represent
the observed distribution, and the solid line the fitted model.
Although our data go as deep at $V \sim 18.0$, we have restricted our 
selection to $V \le 16.0$, in order to minimize residual magnitude 
equation. All our four estimates (Table 3) agree to within their formal errors, thus indicating
that neither cluster membership, nor color or magnitude-related systematics are
significantly affecting our result. In what follows, we will adopt
as the relative cluster proper motion the value determined from the
Gaussian fit of the proper-motion distribution. 
Combining the zero point for field 191 (Table 2) with the relative cluster motion, we
obtain an absolute proper motion 
$(\mu_{\alpha}^*,\mu_{\delta}) = (3.69\pm0.29, -14.88\pm0.26) $ mas~yr$^{-1}$ for NGC 6397.

We note that in this field, we have also measured 25 LEDA confirmed galaxies (Paturel et al. 2005).
Galaxies' magnitude equation on the photographic plates was treated as described in
Girard et al. (1998).
Eliminating three outliers, we obtain a less-precise zero point of 
$(\mu_{\alpha}^*,\mu_{\delta}) = (1.14\pm1.18, 3.97\pm1.28) $ mas~yr$^{-1}$, where
we used a simple average and the dispersion for this estimate.  This 
gives an absolute proper motion for NGC 6397 of
$(\mu_{\alpha}^*,\mu_{\delta}) = (3.63\pm1.19, -14.06\pm1.29) $ mas~yr$^{-1}$, less well-constrained
but consistent with the more precise absolute motion based on the Table 2 zero point.

\begin{table}[htb]
\caption{NGC 6397: Relative cluster motion}
\begin{tabular}{lrrr}
\tableline
\multicolumn{1}{c}{Sample} & \multicolumn{1}{c}{$\mu_{\alpha}^*$} &
\multicolumn{1}{c}{$\mu_{\delta}$} & \multicolumn{1}{c}{N} \\
& \multicolumn{1}{c}{(mas~yr$^{-1}$)} & \multicolumn{1}{c}{(mas~yr$^{-1}$)} & \\
\tableline
RG  pplots@$10\%$   & $4.66\pm0.17$ &  $-10.04\pm0.14$ & 261 \\ 
BHB pplots@$10\%$   & $4.62\pm0.27$ &  $-9.74\pm0.35$   & 69 \\ 
$P_{\mu}^C > 80\%$ pplots@$10\%$   & $4.80\pm0.14$ &  $-10.11\pm0.14$ & 229 \\ 
$V\le 16$ Gaussian fit   & $4.77\pm0.11$ &  $-10.09\pm0.11$  & 438 \\ 
\tableline
{\bf adopted} & $4.77\pm0.11$ &  $-10.09\pm0.11$  & 438 \\
\tableline
\end{tabular}
\end{table}

\begin{figure*}
\includegraphics[angle=-90,scale=0.65]{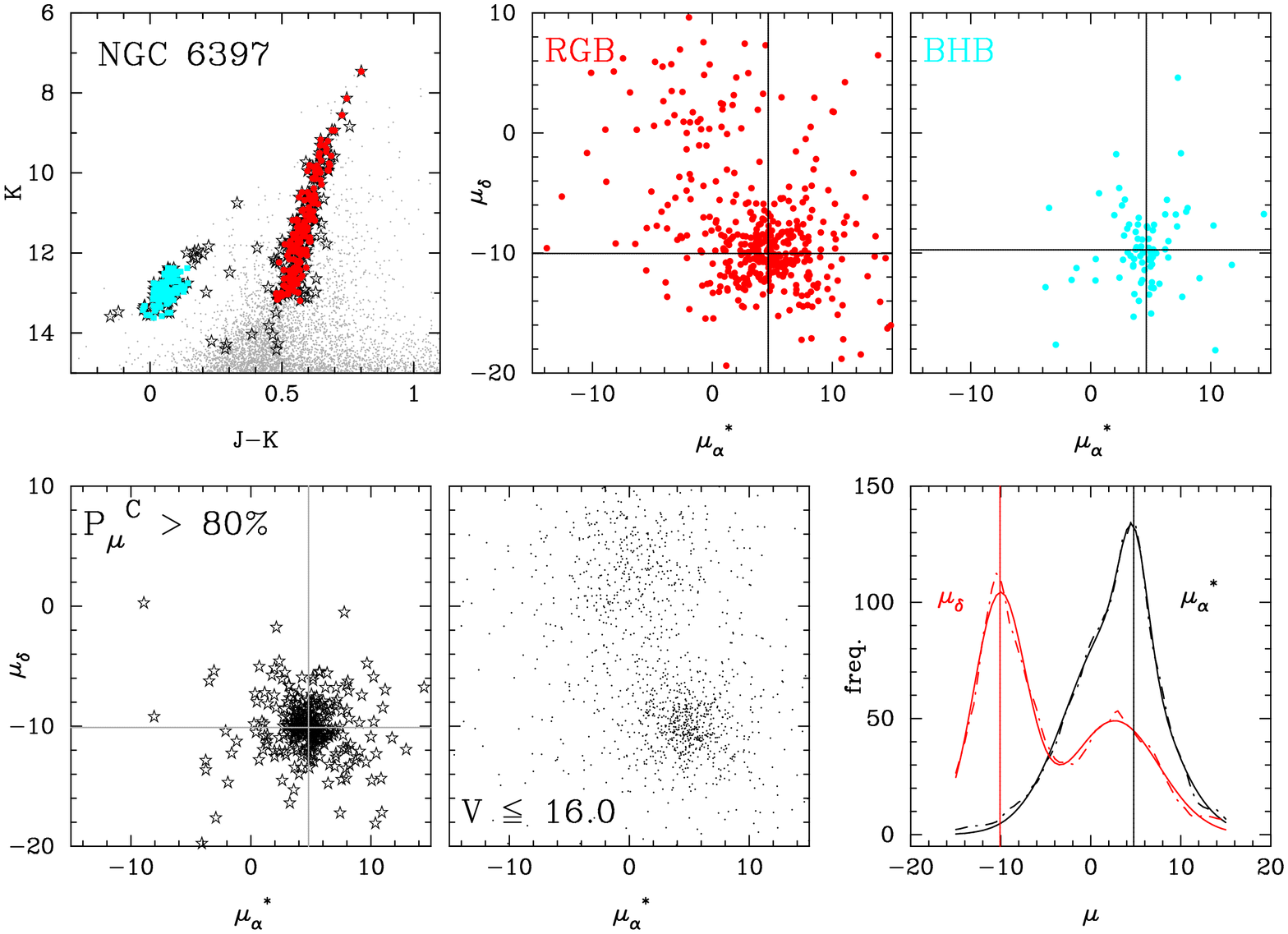}
\caption{2MASS CMD and VPDs of the relative proper motions for NGC 6397.
2MASS-selected red giants are highlighted with red symbols, and blue horizontal branch stars with blue symbols. 
Star symbols show objects with CH93 
proper-motion membership probability larger than $80\%$.
The bottom middle panel shows the VPD of all stars within $11\arcmin$ 
of the cluster center and with $V\le 16$. The observed/fitted
proper-motion distributions for this latter sample are shown with a dashed-dot/solid line in the bottom right panel. 
The mean relative motion for these samples
as derived in the text are also indicated.}
\end{figure*}

For NGC 6626, to separate the cluster from the field we have used only the Rees \& Cudworth (1991) 
proper-motion membership probabilities. In Figure 4, we show the 2MASS CMD 
for stars within $7\arcmin$ of the cluster center.
The RG-selected stars are highlighted in red; also stars with $P_{\mu}^C > 80\%$ are shown with
star symbols. The motion of the cluster is much harder to separate from the field in this case,
as the cluster is more distant and lies closer to
the Galaxy's center than NGC 6397.  The VPDs for the two samples are also shown
in Fig. 4.
The relative proper motion of the cluster is based solely on the
sample with $P_{\mu}^C > 80\%$, for which we derive the mean motion using 
probability plots trimmed at $10\%$. The cross in the right panel of Fig. 4 indicates this value,
which is $(\mu_{\alpha}^*,\mu_{\delta}) = (1.31\pm0.50, -4.75\pm0.58) $ mas~yr$^{-1}$.
Combining with the zero point from Table 2, we obtain the absolute proper motion of NGC 6626 
to be $(\mu_{\alpha}^*,\mu_{\delta}) = (0.63\pm0.67, -8.46\pm0.67) $ mas~yr$^{-1}$.

\begin{figure*}
\includegraphics[angle=-90,scale=0.65]{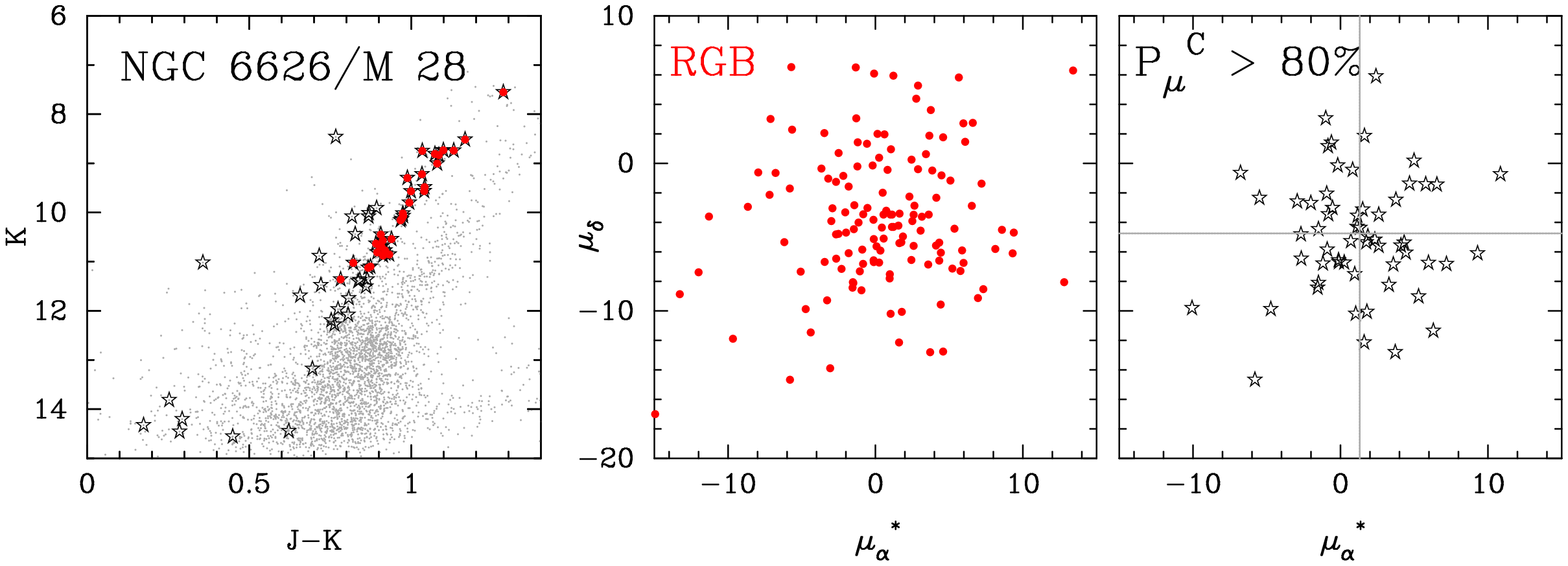}
\caption{As in Fig. 3, but for NGC 6626.
Here we show only the samples of RGs and of stars with Rees \& Cudworth (1991) proper-motion
membership probability larger than $80\%$. The adopted average is from the
latter sample,and is also represented with a cross.}
\end{figure*}

Finally, for NGC 6656, we use all four types of estimates as in the case of NGC 6397.
Proper-motion membership probabilities are from Cudworth (1986), and Peterson \& Cudworth (1994).
Since crowding is severe at the center of the cluster, we use only stars within an annulus 
with  radius $4.6\arcmin \le r \le 11\arcmin$ 
from the cluster center. In Figure 5 we show the 2MASS CMD with 
the respective samples highlighted. The VPDs of RG, BHB, $P_{\mu}^C > 80\%$, and
$V \le 16.0$ stars are shown. The last panel shows the proper-motion distribution
of stars with $V \le 16.0$.
For the $P_{\mu}^C > 80\%$ sample, we estimate the mean using probability plots
trimmed at $10\%$. For the remaining three samples, we apply the
two Gaussian field/cluster decomposition in each proper-motion coordinate.
Since in $\mu_{\alpha}^*$ the separation is better than in $\mu_{\delta}$, we 
first perform the fit in $\mu_{\alpha}^*$, and use the determination of the 
number of cluster stars in this fit as a fixed parameter in the $\mu_{\delta}$ fit.
The estimates of the means are also shown in Fig. 5.
In Table 4 we list these estimates. Errors in the individual
 estimates are done as for NGC 6397.

\begin{table}[htb]
\caption{NGC 6656: Relative cluster motion}
\begin{tabular}{lrrr}
\tableline
\multicolumn{1}{c}{Sample} & \multicolumn{1}{c}{$\mu_{\alpha}^*$} &
\multicolumn{1}{c}{$\mu_{\delta}$} & \multicolumn{1}{c}{N} \\
& \multicolumn{1}{c}{(mas~yr$^{-1}$)} & \multicolumn{1}{c}{(mas~yr$^{-1}$)} & \\
\tableline
RG  Gaussian fit   & $7.57\pm0.15$ &  $-0.10\pm0.19$ & 184 \\ 
BHB  Gaussian fit  & $7.32\pm0.22$ &  $-0.17\pm0.35$   & 43 \\ 
$P_{\mu}^C > 80\%$ pplots@$10\%$   & $6.50\pm0.25$ &  $-0.17\pm0.17$ & 128 \\ 
$V\le 16$ Gaussian fit   & $7.63\pm0.13$ &  $0.39\pm0.09$  & 486 \\ 
\tableline
{\bf adopted} & $7.26\pm0.26$ &  $-0.04\pm0.11$  &  \\ 
\tableline
\end{tabular}
\end{table}

Clearly, the estimate using $P_{\mu}^C$ proper-motion membership probabilities is 
different from the rest in $\mu_{\alpha}^*$. 
We have tried to use different probability cutoffs,
however, the result did not change significantly. 
Likewise, the Gaussian model fit for the $V \le 16$ sample is significantly
different from the other three estimates in $\mu_{\delta}$. 
We suspect that these differences are due to the limitations of the model fits
for small samples and due to residual magnitude equation. 
To this end, we will take 
the average of these four estimates, and its uncertainty is derived from the scatter about the
average. Thus we obtain $(\mu_{\alpha}^*,\mu_{\delta}) = (7.26\pm0.26, -0.04\pm0.11) $ mas~yr$^{-1}$. 
Combining with the zero point from Table 2, the absolute proper motion of NGC 6656 is
$(\mu_{\alpha}^*,\mu_{\delta}) = (7.37\pm0.50, -3.95\pm0.42) $ mas~yr$^{-1}$.

\begin{figure*}
\includegraphics[angle=-90,scale=0.65]{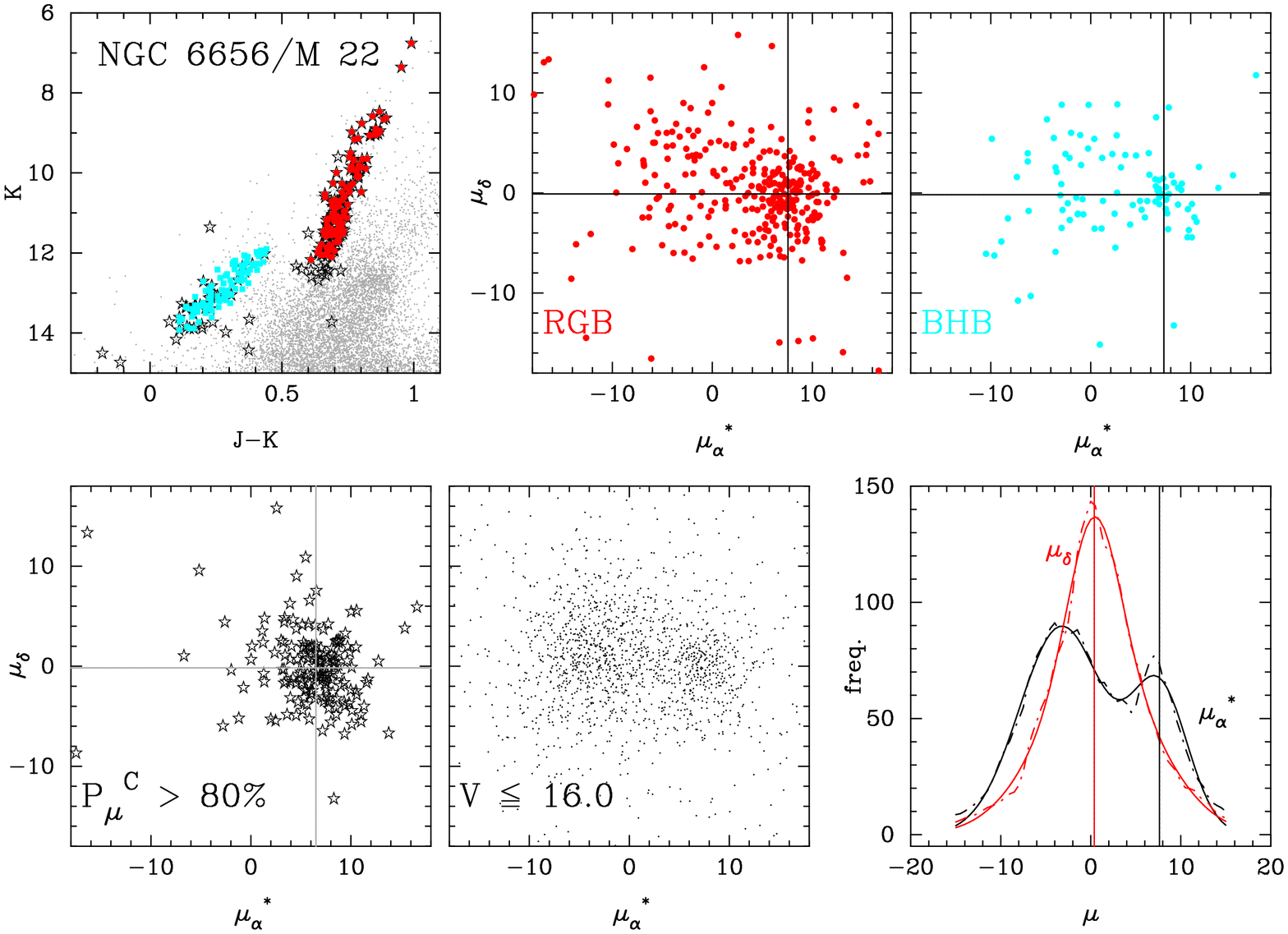}
\caption{As in Figure 3, but for NGC 6656.}
\end{figure*} 

The absolute proper motions of all three clusters are listed in Table 5, together with 
Galactic coordinates, reddening, heliocentric distance and heliocentric line-of-sight velocity.
For NGC 6397, we list the $Hipparcos$-based value.
In Table 6, we list the proper motions along Galactic coordinates, where 
$\mu_l^* = \mu_l$~cos~$b$, as well as the proper motions in a Galactic rest
frame, i.e., with the solar and LSR motion subtracted. The values for the solar and LSR motion 
are given in Section 5. The proper motions in a Galactic rest frame are useful
when displaying the cluster motion on starcount maps of the cluster.


\begin{table*}[tbh]
\caption{Absolute Proper Motions and Other Parameters}
\begin{tabular}{lrrrrrrr}
\tableline
\multicolumn{1}{c}{NGC} &
\multicolumn{1}{c}{$\mu_{\alpha}^*$} & \multicolumn{1}{c}{$\mu_{\delta}$} & 
\multicolumn{1}{c}{$l$} & \multicolumn{1}{c}{$b$} & \multicolumn{1}{c}{$E_{(B-V)}$} & 
\multicolumn{1}{c}{$d_{\odot}$} & \multicolumn{1}{c}{$V_{los}$}  \\
 & \multicolumn{2}{c}{(mas~yr$^{-1}$)} & 
\multicolumn{1}{c}{($\deg$)} & \multicolumn{1}{c}{($\deg$)} & & (kpc) & (km~s$^{-1}$) \\
\tableline
6397 & $3.69(0.29)$ & $-14.88(0.26)$ & 338.17  & -11.96 & 0.18 & 2.3 & $18.8(0.1)$ \\
6626 & $0.63(0.67)$ & $-8.46(0.67)$  &   7.80  & -5.58  & 0.40 & 5.5 & $17.0(1.0)$ \\
6656 & $7.37(0.50)$ & $-3.95(0.42)$  &   9.89  &  -7.55 & 0.34 & 3.2 & $-146.3(0.2)$ \\
\tableline
\end{tabular}
\end{table*}

\begin{table*}[tbh]
\caption{Absolute Proper Motions in Galactic coordinates, and in the Galactic rest frame}
\begin{tabular}{lrrrrrr}
\tableline
\multicolumn{1}{c}{NGC} &
\multicolumn{1}{c}{$\mu_{l}^*$} & \multicolumn{1}{c}{$\mu_{b}$} &
\multicolumn{1}{c}{$(\mu_{l}^*)_{GRF}$} & \multicolumn{1}{c}{$(\mu_{b})_{GRF}$} &
\multicolumn{1}{c}{$(\mu_{\alpha}^*)_{GRF}$} & \multicolumn{1}{c}{$(\mu_{\delta})_{GRF}$} \\
& \multicolumn{2}{c}{(mas~yr$^{-1}$)} & \multicolumn{2}{c}{(mas~yr$^{-1}$)} &
\multicolumn{2}{c}{(mas~yr$^{-1}$)} \\
\tableline
6397 & $-11.06(0.27)$  & $-10.62(0.28)$ & $10.08(0.27)$ & $-11.51(0.28)$  & $15.00(0.29)$ & $3.00(0.26)$ \\
6626 & $-7.22(0.67)$ & $-4.45(0.67)$    & 1.99(0.76) & -4.01(0.67) & 4.48(0.67) & -0.08(0.67) \\
6656 & $-0.26(0.44)$ & $-8.36(0.49)$    & 15.47(0.44) & -7.42(0.49) & 13.52(0.50)  & 10.56(0.42) \\
\tableline
\end{tabular}
\end{table*}

\section{Comparison with Other Studies}


Previous absolute proper-motion determinations for the clusters presented here
have been done.
NGC 6397 has a determination done by CH93, which is based on a model of the
field kinematics to correct for the absolute proper motion. This determination has
been revised by Rees \& Cudworth (2010, and R. Rees, private communication), and
is labeled as RC in our Figure 6. Two other determinations were done using HST images
by Kalirai et al. (2007) and by Milone et al. (2006). The more recent study included 
WFPC2 and ACS images, and used $\sim 400$ galaxies selected by a 
morphological parameter to find galaxies with nearly point-like images for the zero-point determination.
The previous study used only WFPC2 images and some 30 galaxies. Formal errors
of these two studies are 0.04 mas~yr$^{-1}$ for Kalirai et al. (2007) and 0.15 mas~yr$^{-1}$
for Milone et al. (2006). In Figure 6 we show the Rees \& Cudworth, and the 
two HST determinations. We also show our
two determinations: one with respect to $Hipparcos$ stars, the other
with respect to LEDA galaxies. The ground-based results shown here agree within their formal errors.
However, the HST ones do not agree internally, and they both disagree by $\sim 2.5$ mas~yr$^{-1}$
from the ground-based ones. The method used by Rees \& Cudworth and by CH93 to derive the
zero point is based on the Lick Northern Proper Motion Program (NPM1, Klemola et al. 1993) and models the reference stars' proper motions as a
function of apparent magnitude. This can cause problems for low latitude clusters in regions 
of high reddening (Rees, private communication). This was the case for cluster NGC 6121/ M 4, which also
has HST-based, CH93 and SPM-based determinations. In the case of M 4, the SPM result agrees with the HST one,
and differs from the CH93 determination by $\sim 4$ mas~yr$^{-1}$ in $\mu_{\delta}$ (Paper II, 
Bedin et al. 2003,
Kalirai et al. 2004). However, the reddening in the case of M 4 is $E_{(B-V)} = 0.35$, while 
for NGC 6397 it is $E_{(B-V)} = 0.18$ (H96), and therefore the CH93 method may be on a 
safer ground in the case of NGC 6397.  Our ground-based results also reinforce the 
Rees \& Cudworth (2010)
determination. Rees et al. (2013, in prep) and Rees \& Cudworth (2010) also show other 
ground-based determinations
using Tycho2, UCAC3 and PPMXL catalogs. These appear to be in better agreement
with the ground-based determinations shown here, and discrepant from the HST result. 
It is difficult to understand what may have affected the HST result to as much as 0.5 ACS pixels
over $\sim 10$ years in $\mu_{\delta}$.
If the ACS camera is to blame for magnitude equation caused by strong charge transfer inefficiency
(Anderson \& Bedin 2010) then why is the WFPC2-only Milone et al. (2006) result also discrepant
in the same way/amount as the Kalirai et al. (2007) result? 
The HST field used in the Kalirai et al. (2007) study is $5\arcmin$ southeast of the cluster center.
However, cluster rotation is too small to account for the discrepancy seen here ($\sim 0.16$ mas~yr$^{-1}$,
see Kalirai et al. 2007). And, in the Milone et al. (2006) study, three HST fields located around
the cluster give consistent results within their errors. 
Nor can the cluster parallax affect the HST result by  2.5 mas~yr$^{-1}$; the cluster parallax is
$\sim 0.4$ mas projected mostly onto the right ascension.
Unless further analysis of the HST data is done ---
which is beyond the scope of this paper --- we lack a satisfactory explanation for this discrepancy.

\begin{figure}
\includegraphics[angle=-90,scale=0.6]{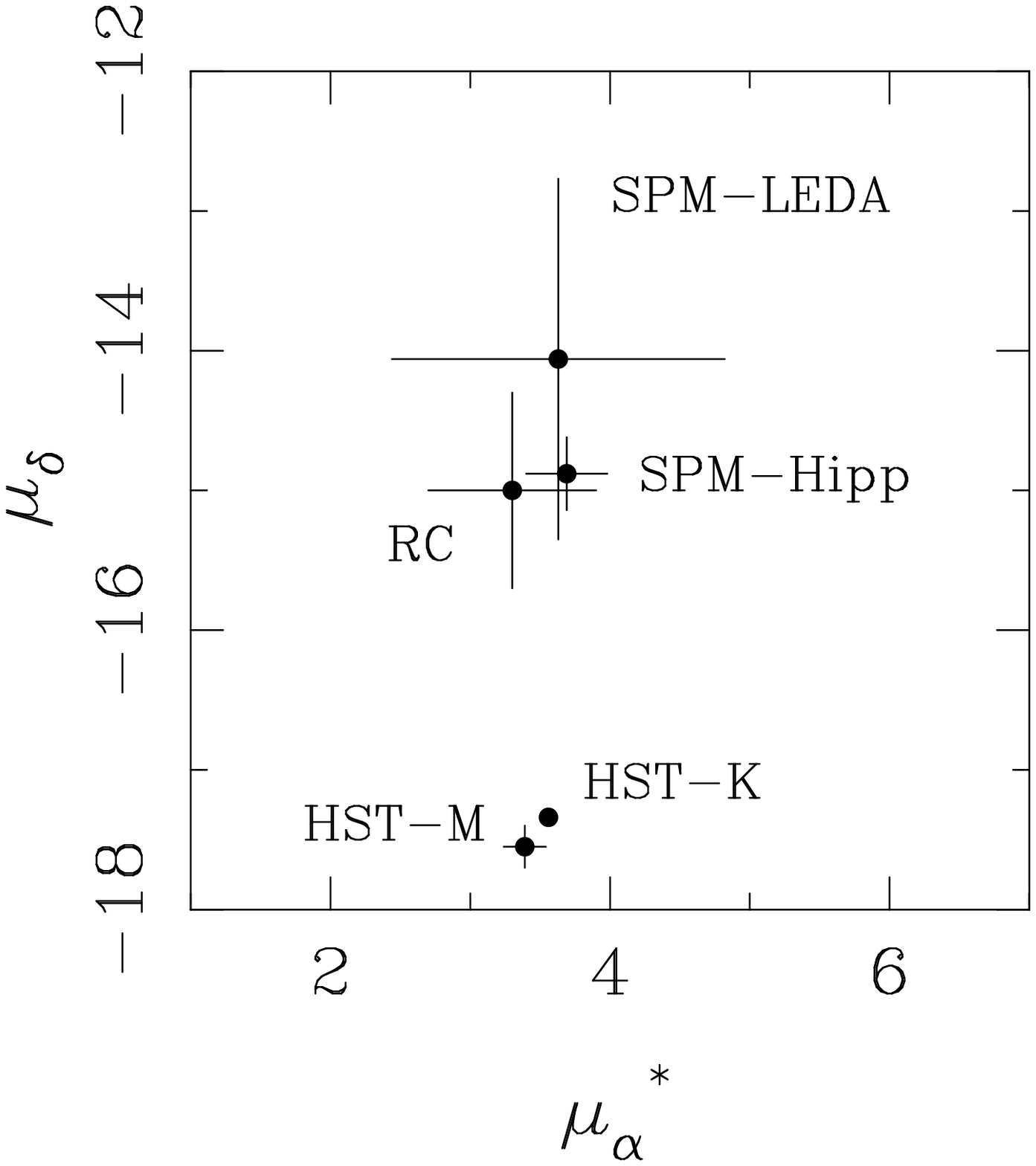}
\caption{Absolute proper-motion determinations for NGC 6397. Ground-based 
determinations agree within their formal errors. However, the HST 
determinations are $\sim 2.5$ mas~yr$^{-1}$ different from the ground-based 
ones in $\mu_{\delta}$.}
\end{figure}


NGC 6626 has been previously measured by CH93. Their result is
$(\mu_{\alpha}^*,\mu_{\delta}) = (0.30\pm0.50, -3.40\pm0.90) $ mas~yr$^{-1}$,
and is discrepant at $4.5\sigma$ level in $\mu_{\delta}$, similar to the case of
M 4. M 28 is also located in a region of high reddening $E_{(B-V)} = 0.40$ (H96), thus
prone to large uncertainty in the predicted kinematics for reference stars.
As shown in the following Section, its newly derived orbit is very eccentric.

Likewise, NGC 6656/ M 22 has been previously measured by CH93, and their determination is
$(\mu_{\alpha}^*,\mu_{\delta}) = (8.6\pm1.3, -5.1\pm1.5) $ mas~yr$^{-1}$, 
in agreement with our measurement. The reddening in this direction is 
$E_{(B-V)} = 0.34$ (H96). Another measurement reported for this cluster is
by Chen et al. (2004) using WFPC2 HST images. The inertial reference in this case
is given by bulge stars. The result is $(\mu_{\alpha}^*,\mu_{\delta}) = (10.19\pm0.20, -3.34\pm0.10)$ mas~yr$^{-1}$,
and discrepant at $5\sigma$ from our result in $\mu_{\alpha}^*$. 
However, we regard this result with some caution as the inertial reference system 
is not ideal. For instance, Pryor et al. (2010) determine the proper motion of Sagittarius dwarf spheroidal (Sgr) core
with respect to foreground stars (in the bulge and thick disk) using HST data. Due to the lack of background galaxies, they have to 
adopt a kinematical model for foreground stars in order to correct Sgr's proper motion to an absolute value, and this correction
is of the order of 6 mas~yr$^{-1}$ in $\mu_{l}$. 
Other studies use background bluge stars in the field of more nearby globular clusters,
assuming bulge stars are at rest; however, this assumption is
applicable for clusters very near to $l = 0\arcdeg$ (e.g., Terndrup et al. 1998, for NGC 6522 at $l=1.03\arcdeg$). NGC 6656 is $\sim 10\arcdeg$ away from $l = 0\arcdeg$. Chen et al. (2004) have not 
considered any type of correction for the motion of reference stars.

\subsection{A Note on Systematics}
 
At the prompting of a referee for this paper, we give below a brief discussion stressing the
distinction between precision and accuracy as it relates to proper-motion measurement,
the practical limitations of ``root N'', and the desire for continued comparison of ground-based
and HST-based measures.

A crucial aspect in determining the accurate {\it absolute} proper motion of a stellar system
is establishing the inertial reference frame, preferably given by extragalactic objects. 
The number of such objects and the precision with which their images can be centered is often 
the limiting factor in such measurements.
For this reason, photographic studies employing plates with excellent scale but relatively
narrow field of view provide precise {\it relative} proper motions and, thus, superb
cluster membership estimates.
However, typically this material is not deep or wide enough to measure sufficient extragalactic 
background objects to yield a useful absolute proper-motion reference frame.
SPM's advantage is that it is deep and wide enough to secure sufficient calibrators
(whether they be $Hipparcos$ stars or galaxies) for reliable absolute proper-motion determination, 
even if the individual proper motions are not as precise as those of studies with better plate scale.
Another advantage of SPM is that magnitude-dependent systematics can be calibrated and removed
due to the multiple images per object (exposures are taken with a diffraction grating) thereby also
allowing the bright reference frame of $Hipparcos$ to be tied to that of the cluster stars.

HST imaging provides extremely precise astrometry over a very small field of view, thus giving
severe importance to background galaxies and the possible systematics
involved in linking them to the target stellar system.
HST-based astrometry of stars versus galaxies was pioneered by J. Anderson and I. King, and further 
developed by J. Anderson (e.g., Anderson \& King 2000, Anderson \& Bedin 2010).
This remains a non-trivial issue given the time dependence of the small but
significant nonlinearity of the HST detectors.
While HST-based errors are formally
very low, it is worth making comparisons to well-established ground-based measures in order to validate
the expected accuracy.
For instance, for both the globular cluster M4 and the Sagittarius dwarf galaxy (Sgr), our 
previously published $Hipparcos$-based results agree within
$1\sigma$ of the subsequent HST measurements,
(M4: Dinescu et al. 1999, Bedin et al. 2003, Kalirai et al. 2004;
Sgr: Dinescu et al. 2005, Pryor et al. 2010). 
Thus, our discrepant result for NGC 6397 came as a bit of a surprise.
It is worth noting that in some cases HST-based absolute proper motions have changed significantly
from an earlier to a later study of the same object.  See, for instance, results for the Fornax 
dwarf spheroidal (Piatek et al. 2002 and Piatek et al. 2007,
a change of $1.7\sigma$), or the more recent cases of the LMC and SMC 
(LMC: Kallivayalil et al 2006a and Kallivayalil et al. 2013 a change of $3\sigma$; 
SMC: Kallivayalil et al 2006b, 2013, a change of $2\sigma$).

\section{Velocities}


Velocities are calculated using the absolute proper motions, distances and heliocentric
line-of-sight velocities 
listed in Table 5. We assume the following: the Sun is at $(x, y, z) = 
(8.0,0.0,0.0)$ kpc,
the Local Standard of Rest (LSR) velocity is 
$(U, V, W) = (0,232,0)$ km s$^{-1}$ (Carlin et al. 2012a), and the Sun peculiar motion is
$(U, V, W) = (-11.10,12.24,7.25)$ km s$^{-1}$ (Sch\"{o}nrich et al. 2010).
Velocity errors are determined from the errors in the proper motions, line-of-sight velocities
and a $10\%$ distance error. In Table 7 we list the velocity components in a Galactic rest frame,
along $(U,V,W)$ directions defined by the location of the LSR,
and in a cylindrical coordinate system along ($\Pi,\Theta$) directions.
$\Pi$ is along the radial direction from the Galactic center to the cluster location 
projected on to the Galactic plane. It is positive outward from the Galactic center. The
$\Theta$ component is perpendicular to the radial direction and positive toward Galactic rotation. 
Along with the velocity components we list
the $(R,z)$ cluster coordinates, where 
$R = \sqrt{(x^2+y^2)}$,  the cluster metallicity and integrated
absolute magnitude from H96. The velocity components indicate prograde, halo-type orbits for all
three clusters. In the following Section we will derive their orbits.

\begin{table*}
\caption{Cluster Positions and Velocities}
\begin{tabular}{rrrrrrrrrr}
\tableline
\multicolumn{1}{c}{NGC} & \multicolumn{1}{c}{[Fe/H]} &
\multicolumn{1}{c}{$M_V$} & \multicolumn{1}{c}{$z$} &\multicolumn{1}{c}{$R$} &
\multicolumn{1}{c}{$U$} & \multicolumn{1}{c}{$V$} & \multicolumn{1}{c}{$W$} & \multicolumn{1}{c}{$\Pi$} &\multicolumn{1}{c}{$\Theta$} \\
 & & & \multicolumn{2}{c}{(kpc)}  &
\multicolumn{3}{c}{(km s$^{-1}$)}   &
\multicolumn{2}{c}{(km s$^{-1}$)}   \\
\tableline
 6397 &-2.02 & -6.64 & -0.5 & 6.0 & 39(06)& 134(11)  & -110(10) & 20(06) & 139(11) \\
 6626 &-1.32 & -8.16 & -0.5 & 2.7 & -42(03) & 58(27) & -110(22) & -24(08) & 68(26) \\
 6656 &-1.70 & -8.50 & -0.4 & 4.9 & 148(02)  & 213(07) & -99(14) & 170(02) & 195(07) \\
\tableline
\end{tabular}
\end{table*}

\section{Orbits}

Orbits were integrated using an axisymmteric gravitational potential and a model including the Galactic bar. The axisymmetric potential is based on the model of Johnston et al. (1995) and its update by Carlin et al. (2012a). It consists of three components: the bulge is represented by a Hernquist potential, the disk by the Miyamoto \& Nagai potential, and the halo by a logarithmic potential. Carlin et al. (2012a) used a triaxial halo (see also Law \& Majewski 2010) and updated the potential of Johnston et al. (1995) to obtain the LSR circular velocity of 232\,km\,s$^{-1}$ at the solar radius of 8\,kpc. We used their updated model parameters except that the halo is taken to be spherically symmetric. For clarity we give the gravitational potential for individual components,
\begin{eqnarray}
\Phi_{\mathrm{bulge}} &=& -\frac{GM_{\mathrm{bulge}}}{r+c},\\
\Phi_{\mathrm{disk}} &=& -\frac{GM_{\mathrm{disk}}}{\sqrt{R^2 + \left( a+\sqrt{z^2+b^2} \right)^2}},\\
\Phi_{\mathrm{halo}} &=& v^2_{\mathrm{halo}} \ln \left(r^2+d^2 \right),
\end{eqnarray}
where $z$ is the vertical coordinate, $R$ is cylindrical radius, and $r$ spherical radius. For the parameters we take: $M_{\mathrm{bulge}}=3.9\times10^{10}$\,M$_{\odot}$, $M_{\mathrm{disk}}=1.1\times10^{11}$\,M$_{\odot}$, $a=6.5$\,kpc, $b=0.26$\,kpc, $c=0.7$\,kpc, $d=12$\,kpc, and $v_\mathrm{halo} = 121.9$\,km\,s$^{-1}$. The axisymmetric potential was also used in the model including the Galactic bar, where, however, the bulge mass $M_\mathrm{bulge}$ is reduced by the mass of the bar.

The potential of the Galactic bar is adopted from J\'{i}lkov\'{a} et al. (2012), where the bar is modeled by a Ferrers potential of an inhomogeneous triaxial ellipsoid (Binney \& Tremaine 2008, where we choose $n=2$). Parameters of the bar potential are based on Pichardo et al. (2004): the bar  mass is $0.98\times10^{10}$\,M$_{\odot}$ ($M_\mathrm{bulge}$ is decreased by this mass), it rotates with angular velocity $60$\,km\,s$^{-1}$\,kpc$^{-1}$, and the semi-axes lengths are 3.14, 1.178, and 0.81\,kpc (see Table\,2 of J\'{i}lkov\'{a} et al. 2012). Such a model together with the axisymmetric background described above give a bar strength parameter of 0.27 (maximal value of the ratio between the tangential force due to the non-monopole part of the potential and the radial force due to the monopole part at the same radius).

The orbits were integrated using a Bulirsch--Stoer integrator with adaptive time-step (Press et al. 1992). The relative change of the total energy in the axisymmetric case over an integration time of 5\,Gyr is of the order of 10$^{-15}$, and the relative change of the Jacobi energy in the barred case is 10$^{-13}$. In Table~8 we list the orbital parameters as averages over the number of cycles in a 5\,Gyr 
integration back in time.
The uncertainties in the orbital parameters are derived from the dispersion in each element obtained over 1000 repeats of the initial conditions for each cluster. Initial conditions are obtained from Gaussian distribution values drawn for the proper motions, distance and line-of-sight velocity according to the errors of these quantities. For each cluster, the first line shows the parameters for the
axisymmetric model (a), while the second line for the bar model (b). 
Minimum and maximum orbital angular momentum $L_z$ values are given for the bar model.

For NGC 6397, we also list the orbital parameters for 
the HST proper-motion determination, where we took the average of the Milone et al. (2006) and the Kalirai et al. (2007)  
proper-motion determinations, and the remaining parameters from Table 5 and initial conditions.
The HST measurement gives a more disruptive orbit than the ground-based
result, due to a smaller pericentric radius, and more frequent pericenter and disk passages
due to a shorter period orbit; however orbital parameters differ only at $\sim 1.4\sigma$ level.
For each cluster, orbital parameters as given by both models are similar within the observational uncertainties.

We note that the orbit of NGC 6626 is irregular, and --- in the barred model --- $L_z$ changes its sign in many of the
1000 realizations of the orbit integration.

\begin{table*}\small
\caption{Orbital Parameters}
\begin{tabular}{lrrrrrcc}
\tableline
\multicolumn{1}{c}{NGC} & 
\multicolumn{1}{c}{$P_{azi}$} & 
\multicolumn{1}{c}{$r_a$} & 
\multicolumn{1}{c}{$r_p$} &
\multicolumn{1}{c}{$|z_{max}|$} & 
\multicolumn{1}{c}{ecc} & 
\multicolumn{1}{c}{$E_{orb}$, $E_{\mathrm{J}}$} & 
\multicolumn{1}{c}{$L_z^{\ddagger}$} \\
& \multicolumn{1}{c}{(Myr)} & \multicolumn{1}{c}{(kpc)} & \multicolumn{1}{c}{(kpc)} & \multicolumn{1}{c}{(kpc)} & &
\multicolumn{1}{c}{($10^2$ km$^2$~s$^{-2}$)} & \multicolumn{1}{c}{(kpc~km~s$^{-1}$)}  \\
\tableline
6397 a              & 132.8(04.3) &  5.97(0.21) &  2.91(0.26) &  1.76(0.26) &  0.34(0.03)  &   153(13)  &    $-$829(070)  \\
6397 b              & 123.7(10.2) &  5.89(0.48) &  2.71(0.39) &  1.73(0.25) &  0.37(0.05)  &   $-$353(30) &   $-$837(100)... $-$789(111)   \\ 
6397$^{\dagger}$ a  & 128.8(03.2) &  6.00(0.18) &  2.34(0.31) &  2.11(0.26) &  0.44(0.05)  &   145(08)  &    $-$705(074)   \\
6397$^{\dagger}$ b  & 114.3(10.7) &  5.66(0.57) &  2.06(0.47) &  2.03(0.32) &  0.47(0.09)  &   $-$287(36) &  $-$683(114)... $-$624(121)   \\ \\
6626 a              &  50.6(10.2) &  2.60(0.58) &  0.55(0.27) &  1.33(0.30) &  0.66(0.09)  &   $-$295(88) &    $-$194(093)   \\
6626 b              &  51.0(09.8) &  2.52(0.48) &  0.59(0.26) &  1.19(0.28) &  0.64(0.11)  &   $-$471(65) &  $-$288(092)...  $-$177(089)  \\ \\
6656 a              & 169.3(07.9) &  8.76(0.42) &  2.76(0.24) &  1.49(0.30) &  0.52(0.02)  &    279(24) &    $-$956(079)   \\
6656 b              & 158.0(11.7) &  8.50(0.65) &  2.64(0.26) &  1.45(0.35) &  0.53(0.02)  &   $-$309(30) &  $-$970(090)... $-$920(093)  \\
\tableline
\multicolumn{8}{l}{$\dagger$ HST proper-motion determination (see text).} \\
\multicolumn{8}{l}{$\ddagger$ $L_z^{LSR} = -1856$ kpc~km~s$^{-1}$. For the bar model, we list minimum and maximum values.} \\
\end{tabular}
\end{table*} 

In Figures 7 and 8 we show the orbits in the axisymmetric and bar model, respectively, based on the observed parameters in Table 5.
Each row of panels shows a cluster (for NGC 6397, we use our current proper-motion determination);
the orbits are shown in the Galactic plane, and in cylindrical coordinates in the $(R,z)$
plane. The Sun's location and the LSR velocity is also indicated. 
Orbits for all clusters are plotted on the same scale.
The most confined orbit is that of NGC 6626, while that of NGC 6656 is the most energetic.
Both NGC 6397 and NGC 6626 are now at apocenter, while NGC 6656 is at
4.9 kpc from the Galactic center, traveling toward apocenter. Orbits are prograde (except for some realizations for NGC 6266 in the bar model
where $L_z$ changes sign), and within $\sim 1.5$ kpc from the Galactic plane. All three clusters had recent
crossings of the Galactic plane some 4 Myr ago.

\begin{figure*}
\includegraphics[scale=0.9]{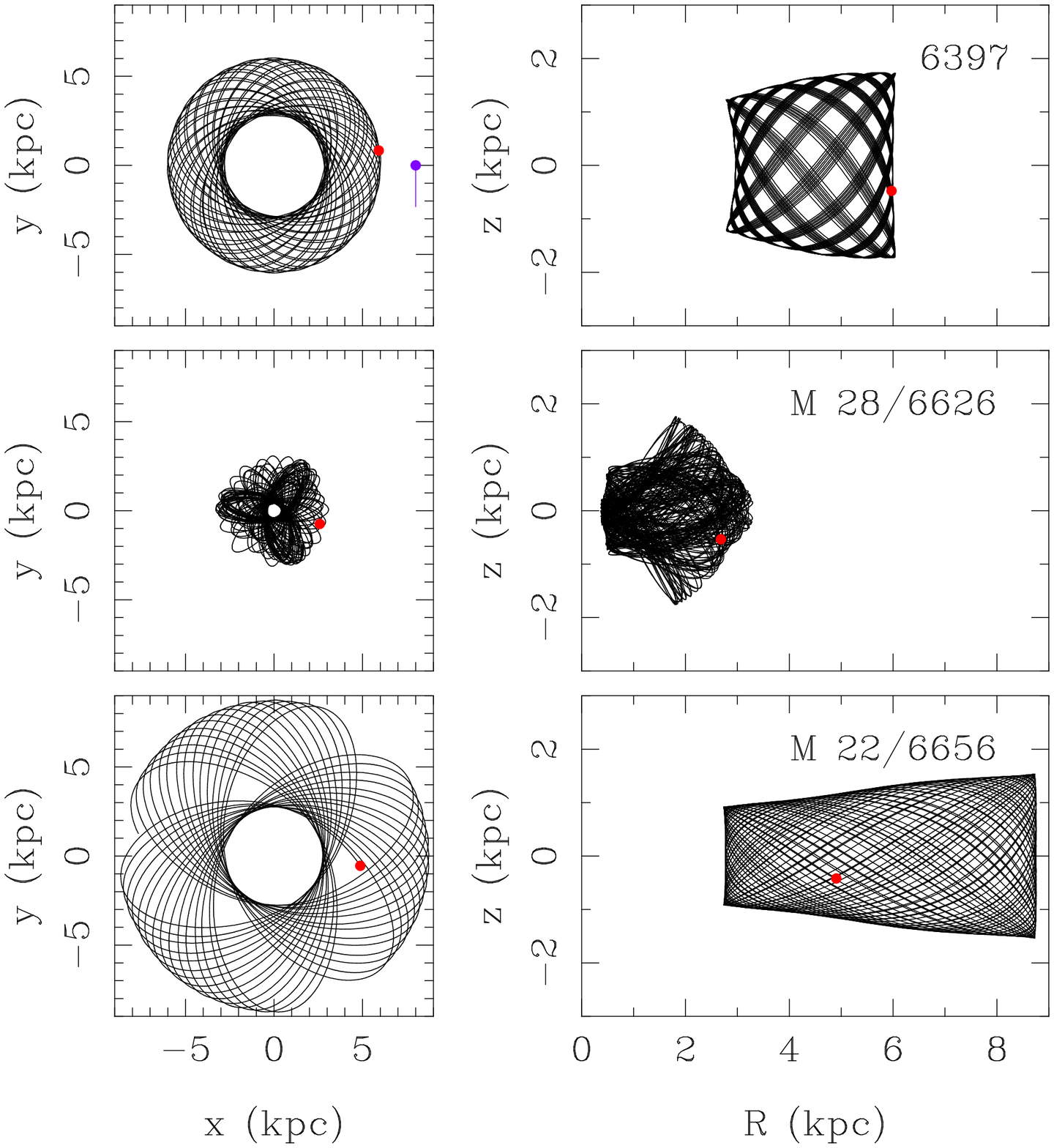}
\caption{Orbits of the three clusters in the axisymmetric model. The location of the Sun, and the 
LSR velocity are indicated in the top-left panel.
The current location of each cluster is shown with a red symbol.}
\end{figure*}
\begin{figure*}
\includegraphics[scale=0.9]{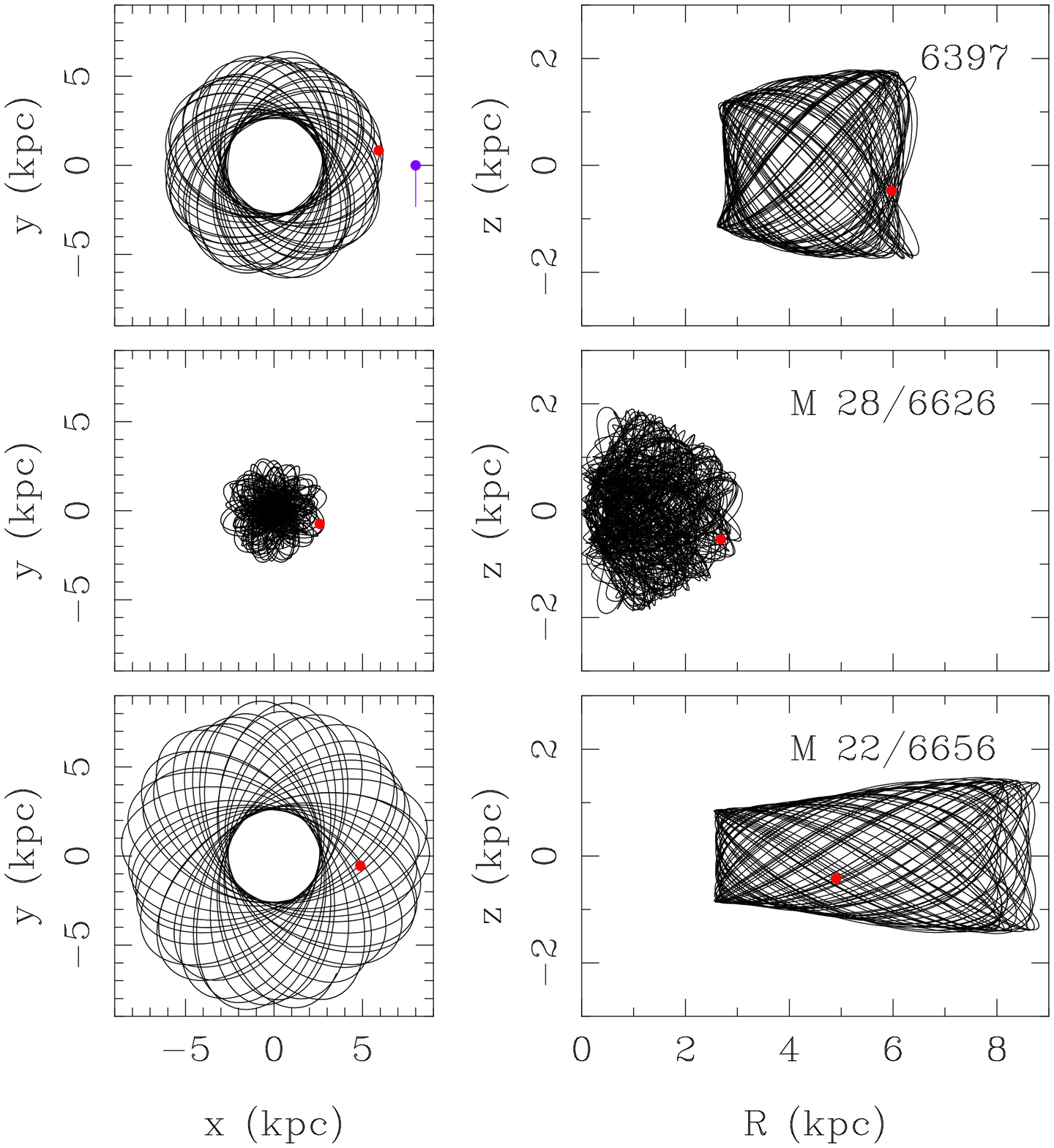}
\caption{As in Figure 7, but for the bar model.}
\end{figure*}

\section{Discussion}

\subsection{NGC 6397}

NGC 6397 is a low mass, very metal-poor, and compact (core collapsed) cluster (see, e.g. H96),
which is now known to display a double main sequence (Milone et al. 2012) due to
two different populations: one similar to field stars, the other enhanced in light elements
such as sodium and nitrogen and depleted in carbon and oxygen. Its orbit is a typical halo-like one,
confined to the inner regions of the Galaxy, and thus prone to substantial disk and bulge shocking along its 
life.

\subsection{NGC 6626/ M 28}

This cluster is a massive, moderately metal poor system (see Table 5), with a moderately extended blue horizontal branch
(according to the Lee et al. 2007 classification). Its orbit is rather disruptive, indicating that this system may have
lost substantial mass. In the recent work by Chun et al. (2012), where the cluster was imaged with a wide-field
near-infrared camera on the CFHT to a limiting $K \sim 19$, an extended stellar feature is uncovered in the 
CMD-filtered density map of the cluster. This feature extends toward the west and slightly north
 direction (see their Fig. 5), from within the tidal radius and beyond it, limited by the area coverage of 
the study ($21\arcmin\times21\arcmin$). Interestingly, our proper-motion vector (with the Solar motion 
subtracted, see Table 6), is well aligned with this stellar extension, and the sense of motion is 
opposite to the stellar extension, thus suggesting a trailing tidal tail of the cluster. 
N-body simulations of clusters moving through a MW potential by Montuori et al. (2007) show that
clusters on very eccentric orbits at pericenter may develop inner (within a few tidal radii)
 tails that correlate with the orbit of the cluster,
while for the rest of the orbit, the inner tails are oriented toward the Galactic center direction.
While NGC 6626 has a rather eccentric orbit, its location is at apocenter. A second stellar extension
less pronounced than the first one is seen in the direction toward the Galactic center in Fig. 5 of
Chun et al. (2012). We also note that the direction perpendicular to the Galactic plane
is also toward northwest, and since the cluster had a recent ($\sim 4$ Myr ago) plane crossing, 
it is possible that the most prominent tidal extension toward northwest is due to this passage. 

\subsection{NGC 6656/ M 22}

M 22 is another relatively massive, metal poor cluster with a well known extended horizontal branch (e.g., Lee et al. 2007
and references therein, and Joo \& Lee 2013). 
It is among the six clusters (according to the review by Gratton et al. 2012) with a spread in 
iron abundance. M 22 has an iron spread of about 0.15 dex (Da Costa et al. 2009, Marino et al. 2009, 2011), 
and HST photometry reveals that it has a bimodal subgiant branch (Piotto et al. 2012) which correlates with the 
different abundances (Marino et al. 2012), in the sense that the fainter subgiant branch has 
stars with higher metallicity. 
The currently proposed scenario for clusters with clear iron abundance spread is that their progenitors were
the most massive of all the progenitors of globular clusters such that they were able to retain 
ejecta from core-collapsed SNe, and form a next generation of stars from this enriched environment.
The emerging picture from both photometry and spectroscopy is that most globular clusters
host two (or more) distinct stellar populations (e.g., Gratton et al. 2012
and references therein): the first has abundance patterns similar to halo stars and is in proportion of
$\sim 30\%$, while the second displays the Na, N, Al, He-enhanced and C, O, Ne, Mg-depleted pattern, and is in proportion of $\sim 70\%$.
The former is interpreted as the first generation of stars, while the latter as the second generation
formed from a combination of enriched material by the first generation and some pristine
material. The details of the enrichment of the second generation (and of following ones) appear to be regulated
by the ability of the progenitor to retain ejecta, and this has to do primarily with the initial mass
of the progenitor (e.g., Valcarce \& Catelan 2011, 
Gratton et al. 2012). That the fraction of second-generation stars is substantially larger than that of the
first-generation in present-day clusters, implies abundant mass loss from the initial mass of the progenitor.
Some authors estimate that the progenitor mass was a factor of $10 - 20$ larger than that of the present-day cluster mass
(Conroy 2012, Carretta et al. 2010, D'Ercole et al. 2008), while others indicate that the mass loss was as much as $99\%$ of the progenitor mass
(Valcarce \& Catelan 2011). Based on the complex abundance patterns and
multiple sequence CMDs of
NGC 2808 and $\omega$ Cen, Valcarce \& Catelan (2011) estimate progenitor masses of 
$ 3.8 - 5.9 \times 10^{8} M_{\odot}$ and $2.2 - 7.4 \times 10^{9} M_{\odot}$ respectively.
Interestingly, an N-body calculation of the disruption and orbital decay of $\omega$ Cen
by Tsuchiya et al. (2003) also suggests a massive progenitor of $ 8 \times 10^{9} M_{\odot}$, in order to obtain the present-day
mass and Galactic location of the cluster. The progenitor in the 
Tsuchiya et al. (2003) study was launched from 58 kpc from the Galactic center, and
had a shallow density profile with a nuclear concentration that survived the tidal disruption to the present day.

Thus, with an iron abundance spread, M 22 likely also had
a very massive progenitor, larger than that of NGC 2808 (which has no iron spread, but three distinct main sequences),
however, smaller than that of $\omega$ Cen with an iron abundance spread of 1.5 dex. 
M 22's orbit is very different from that of $\omega$ Cen (which is retrograde and more eccentric).
The other clusters with confirmed iron abundance spread are M 54 (at the core of Sagittarius dwarf galaxy - Sgr, and likely its nucleus),
Terzan 5, and possibly NGC 1851 and NGC 2419 (Gratton et al. 2012 and references therein). 
NGC 1851's iron spread was questioned by Villanova et al. (2010), while NGC 2419 shows a spread in Ca but not in [Fe/H] (Cohen \& Kirby 2013).
Still, among the most massive clusters of our Galaxy, NGC 2419 is believed to have been part of a much larger system. Its likely association with
the Virgo stellar overdensity (Carlin et al.~2012b, Casetti-Dinescu et al.~2009) reinforces this picture.

Of these, only NGC 1851 has an orbit determination (Paper I), while M 54 can be
assigned a similar orbit to that of the core of Sgr (measured by Dinescu et al.~2005, Pryor et al.~2010) if assumed its nuclear cluster.
The orbits of NGC 1851 and of M 54 are unrelated to each other (although prograde, they are of different orbital inclination) and also unrelated to that
of $\omega$ Cen. Likewise, M 54 and M 22 have different orbits: M 54's is nearly polar like Sgr's orbit, while M 22 is confined to the 
Galactic plane. A possible dynamical link between NGC 1851 and M 22 may exist as the orbits have similar eccentricities and
inclinations, however only if one assumes significat orbital decay for M 22, 
as their orbital energies differ substantially.
NGC 2419 is located in the outskirts of the Galaxy at a galactocentric distance of 90 kpc, while Terzan 5 is within 1.2 kpc
of the Galactic center (H96). Therefore, from the available orbital information and from the current Galactic location of these
six clusters, it appears that they are dynamically unrelated, and each formed in its own parent system.
Therefore, we are seeing at least six 
(if no more iron-spread clusters are found) massive progenitors that hosted
the formation of these chemically complex clusters before they were completely destroyed by internal and external (orbital) dynamical forces; 
five if NGC 1851 is excluded.
We also note that these six clusters are spread over a large Galactocentric distance, 
with apocenter radii of 6, 8.5, 33, and 52 kpc for $\omega$ Cen, M 22, NGC 1851, and M 54, respectively. 
The most distant ones from the Galactic center, M 54 in the Sgr system and NGC 2419 are likely recent accretion events.

If the masses of these progenitors are indeed of the order of
$10^{8} - 10^{9} M_{\odot}$, then these can be thought of as the handful of primordial building blocks of the Milky Way halo 
predicted by cosmological simulations such as those described by Cooper et al. (2010). 
Cooper et al. (2010) predict that MW-type halos are assembled from at most five massive subhalos between
redshift $ 1 < z < 7$; however their subhalo masses are of the order of $10^{7} - 10^{8} M_{\odot}$.
Therefore, better determinations of the primordial masses of the progenitors of iron-spread globular clusters are
needed. If these systems were indeed as massive as predicted by e.g., Valcarce \& Catelan (2011), they may have also hosted 
the formation of a few other less-massive clusters, with the central, nuclear one, being the most chemically-enriched.

We would like to thank Rick Rees and Kyle Cudworth for providing us their 
data for the clusters and for many extensive and valuable discussions.
We acknowledge support by the NSF through grant AST04-0908996.
This publication makes use of data products from the Two Micron All Sky Survey, 
which is a joint project of the University of Massachusetts and the Infrared
Processing and Analysis Center/California Institute of Technology, funded 
by  NASA and  NSF.

\end{document}